# Hartmann vs. reverse Hartmann test: a Fourier optics point of view

François Hénault, Cyril Pannetier
Institut de Planétologie et d'Astrophysique de Grenoble
Université Grenoble-Alpes, Centre National de la Recherche Scientifique
B.P. 53, 38041 Grenoble – France

## ABSTRACT

The Shack-Hartmann Wavefront Sensor (WFS) is well-known in the fields of optical metrology, wavefront sensing in astronomy, and ophthalmologic control applications. The purpose of this communication is to bring new insights on the historical Hartmann test and to compare it with the less known reverse Hartmann test, where the locations of the pupil mask and observed image planes are exchanged. Both tests can actually be interpreted by using the formalism of Fourier optics, i.e. Fraunhofer diffraction for the Shack-Hartmann and Fresnel diffraction in the reverse configuration. The principles of these models are firstly described in the communication. The results of numerical simulations are then presented, allowing comparing both optical arrangements from the Fourier optics point of view, in terms of achievable wavefront measurement accuracy. They show that a WFS based on the reverse Hartmann test may globally achieve the same performance as the classical Shack-Hartmann.

## 1 INTRODUCTION

The Shack-Hartmann (SH) Wavefront Sensor (WFS) is well-known in the fields of optical metrology [1], wavefront sensing in astronomy [2], and ophthalmologic control applications [3]. Although SH-WFS are the subject of extensive literature, it is most generally presented and discussed in the theoretical frame of geometrical optics (with some noticeable exceptions, see e.g. Ref. [4]). One primary goal of this communication is to present an alternative view of the SH-WFS by using the formalism of Fourier optics. For that purpose, it is helpful to compare the classical SH concept, itself based on the historical Hartmann test, with the less known "reverse Hartmann" test configuration, where the locations of the pupil mask and observed image planes are exchanged. This reverse test has recently been described with the help of Fresnel diffraction theory [5], thus paving the way for a new generation of Reverse Hartmann (RH) WFS. In that case the reconstruction procedure of the slopes of the Wavefront Error (WFE) only makes use of Fourier analysis. It will be shown that the same algorithms can be applied to the SH-WFS, giving birth to a new data processing scheme named SH-IFT (Shack-Hartmann inverse Fourier transform) and providing a basis for comparison between the SH-WFS and RH-WFS.

For the sake of completeness, the principles of the RH-WFS are firstly summarized in section 2. The application of the principle to the SH-WFS is then described in section 3. Both optical arrangements are discussed and illustrated with the help of numerical simulations in section 4, allowing comparing them in terms of achievable WFE measurement accuracy. Concluding remarks are given in section 5.

# 2    A FOURIER OPTICS VIEW OF THE REVERSE HARTMANN TEST

In this section is firstly presented the principle of the reverse Hartmann test (§ 2.1). Its Fourier optics theory is summarized in § 2.2. The WFE slopes reconstruction procedure is described in § 2.3. System optimization in view of building efficient and operational RH-WFS is finally discussed in subsection 2.4.

## 2.1    Principle

An illustrative presentation of the reverse Hartmann test and the way it differs from the classical "direct" test is shown in Figure 1. In the direct test (see Figure 1-A) a grid of equally-spaced pinholes is set at the exit pupil plane XY of the optical system to be measured, e.g. a telescope mirror. The observation plane X'Y' is located near (but not at) the focal plane of the system, where a small distorted replica of the pupil grid is recorded either by a photographic plate or a modern camera. Provided that the distance from focus to X'Y' plane is large enough, each observed spot in the X'Y' plane can be associated to a single pinhole source in the pupil plane. Then the deviations with respect to a perfect replica of the pupil grid are directly proportional to WFE slopes errors in the XY plane. Modern SH-WFS are still operating from that basic and classical principle.

On the contrary, the reverse Hartmann test consists in inverting the locations of the grid and its image. The pinhole grid is set near (but not at) the X'Y' plane and its distorted image is observed at the XY pupil plane by means of a backward gazing camera. One intuitively realizes that the same information about WFE slopes errors can be derived from the distortions of the observed grid in the pupil plane. In the two next subsections are summarized the Fourier optics model of the images observed by the backward gazing camera, and the digital procedure employed for WFE slopes reconstruction.

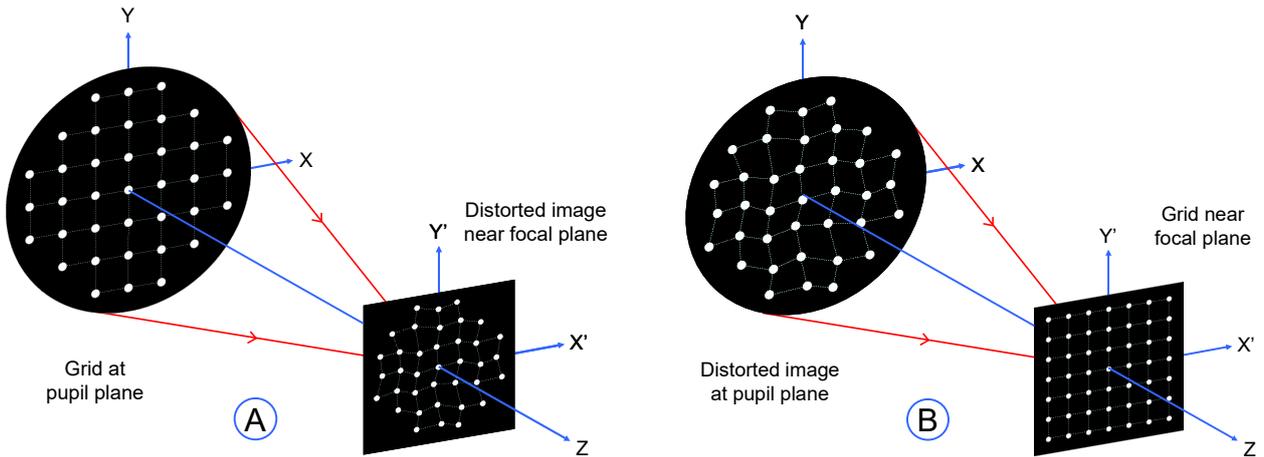

Figure 1: Illustrating the direct Hartmann test (A) and its reverse version (B).

## 2.2    Fresnel diffraction theory of reverse Hartmann test

A complete theoretical description of the reverse Hartmann test cannot be achieved without using Fresnel diffraction formalism, as demonstrated in Ref. [5]. The general optical configuration of the reverse Hartmann test is depicted in Figure 2, showing the main coordinate systems and employed parameters:

- The exit pupil of the tested optical system is located in the OXY plane.

- The Hartmann grid is located at the so-called "spatial filter" plane O'X'Y'.

- The backward gazing camera is located at the focal plane $O_IUV$ of the optical system, and forms an image of the pupil plane on a detector array. Core parameters $F$ and $z'$ respectively denote the focal length of the system and the distance $O'O_I$ from the focal plane to the filter.

Moreover the type and transmission map of the employed spatial filter are of prime importance [5]. Two different types of filter are considered here, as illustrated in Figure 3:

1) The Hartmann mask that is a grid of equally-spaced pinholes. Its transmission function $T_{RH}(x',y')$ is written analytically as:

$$T_{RH}(x',y') = \sum_{m=-M}^{+M} \sum_{n=-M}^{+M} B_{d'}(x'-mp', y'-np') \qquad (1)$$

for a number $N = 2M+1$ of pinholes along both X' and Y' axes, and where $d'$ is their common diameter, $p'$ their spacing assumed to be identical along both axes, and $B_{d'}(x',y')$ stands for a "pillbox" function equal to unity inside a circle of diameter $d'$ and to zero outside of it.

2) The square Ronchi grid that may be seen as a spatially continuous version of the previous one, writing as:

$$T_{SR}(x',y') = [1 + \cos(2\pi x'/p')][1 + \cos(2\pi y'/p')]/4 \qquad (2)$$

Figure 3 also shows false-colour views of the power spectrum of these filters. One notes the presence of numerous harmonic peaks in the spectrum of the Hartmann mask, which do not appear with the square Ronchi grid. This suggests that the latter carries useful information more efficiently than the former. Moreover, Eq. 2 is quite simpler and well-suited to an analytical development than Eq. 1. Thus for the case of a square Ronchi grid filter and provided that system parameters $z'$ and $p'$ are properly sized (see § 2.4), the analytical expression of the intensity distribution $I(x,y)$ observed at the pupil plane – sometimes named "Hartmanngram" – is found to be [5]:

$$I(x,y) = B_D(x;y) \left\{ \frac{3}{8} + \frac{1}{4} C(\lambda) \cos\left[ G\left( \frac{\partial \Delta(x,y)}{\partial x} + \frac{x}{d''} \right) \right] + \frac{1}{4} C(\lambda) \cos\left[ G\left( \frac{\partial \Delta(x,y)}{\partial y} + \frac{y}{d''} \right) \right] \right.$$
$$\left. + \frac{1}{16} \cos\left[ 2G\left( \frac{\partial \Delta(x,y)}{\partial x} + \frac{x}{d''} \right) \right] + \frac{1}{16} \cos\left[ 2G\left( \frac{\partial \Delta(x,y)}{\partial y} + \frac{y}{d''} \right) \right] \right\} \qquad (3a)$$

where $B_D(x;y)$ is the pupil mask function, $\partial \Delta(x,y)/\partial x$ and $\partial \Delta(x,y)/\partial y$ are the slopes of the wavefront $\Delta(x,y)$ to be measured, and the following parameters are used.

Gain factor: $\qquad G = 2\pi(F+z')/p'$, $\qquad$ (3b)

Fresnel diffraction parameter: $\qquad d'' = F(F+z')/z'$, $\qquad$ (3c)

Contrast factor of the Hartmanngram: $\qquad C(\lambda) = \cos\left[ \pi \frac{\lambda(F+z')^2}{d''p'^2} \right]$, $\qquad$ (3d)

with $\lambda$ the wavelength of the incoming radiation.

It must be noted that the equivalent expression of $I(x,y)$ for the case of the Hartmann mask should be much more complicated. However it shares the common property with Eqs. 3 that the locations of maximal peak intensities are found at the same points of Cartesian coordinates $(x_m, y_n)$ in the XY pupil plane, i.e.

$$\begin{aligned} x_m &= mF\,p'/z' - d''\,\partial \Delta(x,y)/\partial x \\ y_n &= nF\,p'/z' - d''\,\partial \Delta(x,y)/\partial y \end{aligned}, \qquad (4)$$

where the reader will recognize a clear similarity with the spot location deviations measured on the detector array of a SH-WFS (see § 3.1).

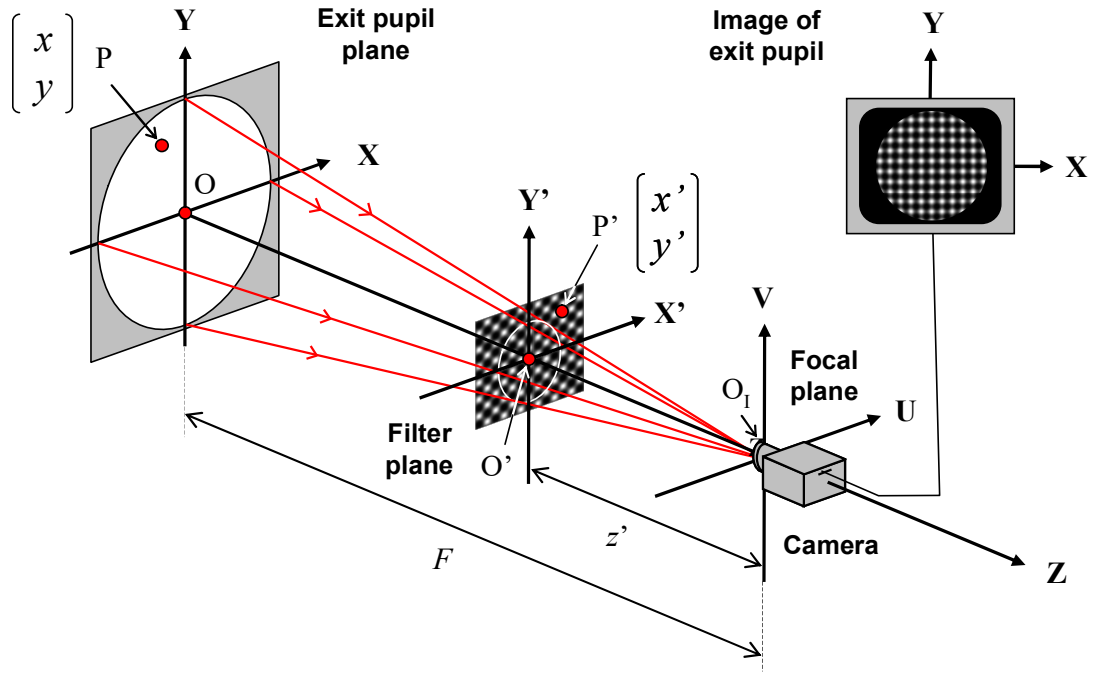

Figure 2: General optical configuration of the reverse Hartmann test.

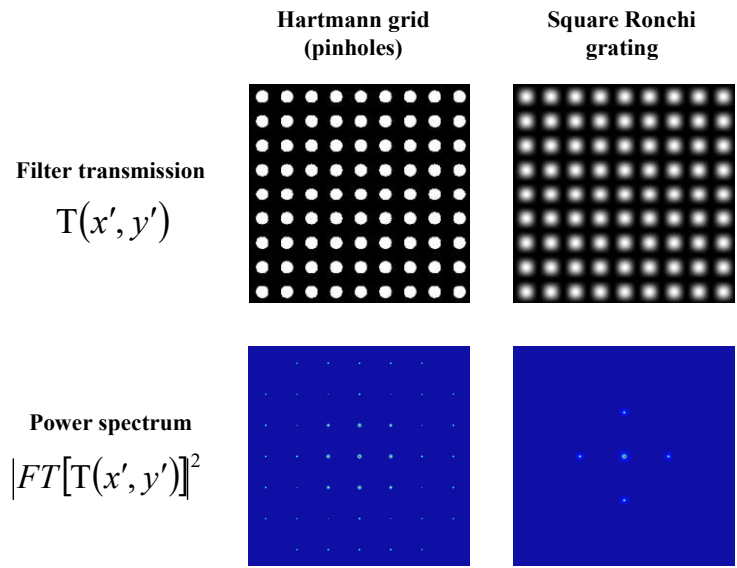

Figure 3: Hartmann mask (left), square Ronchi grid (right), and their power spectrums (lower part of the figure).

## 2.3 Reconstructing WFE slopes

The WFE slopes reconstruction procedure applied to the reverse Hartmann test is inspired from the double Fourier transform algorithm firstly described by Takeda *et al* [6]. Mathematically, it writes as:

$$\frac{\partial \Delta(x,y)}{\partial x} = \frac{1}{G} Arg\left\{FT^{-1}\left[B_{z'/p'F}(u,v) FT[I(x,y)](u-z'/p'F, v)\right]\right\}$$
$$\frac{\partial \Delta(x,y)}{\partial y} = \frac{1}{G} Arg\left\{FT^{-1}\left[B_{z'/p'F}(u,v) FT[I(x,y)](u, v-z'/p'F)\right]\right\}$$
(5)

where $FT[-]$ and $FT^{-1}[-]$ denote direct and inverse Fourier transforms respectively, and $Arg[-]$ is the argument of a complex number. The procedure is also illustrated in Figure 4 and summarized below (same numbers as appearing in the figure).

1) The first step consists in recording the image $I(x,y)$ of the exit pupil seen through the spatial filter or the Hartmann mask, from the backward gazing camera. As expected, it looks as a distorted replica of the real mask.

2) The Fourier transform of the intensity distribution $I(x,y)$ is computed. The result shows a series of regularly-spaced harmonic peaks. The number of peaks depends on the nature of the employed spatial filter (numerical simulations in Ref. [5] showed that more peaks are observable with the Hartmann mask).

3) Whatever is the type of the mask, the first side lobe located in the UV Fourier plane at a distance $z'/p'F$ from the origin along the U-axis is isolated and reentered on the origin.

4-5) The inverse Fourier transform of the recentred side lobe is then computed. The result is a complex function whose modulus (4) is a blurred version of the pupil function $B_D(x;y)$ and phase (5) is directly proportional to the WFE slopes along X-axis $\partial \Delta(x,y)/\partial x$.

6) The same operation than in step 3) is applied to the first side lobe of the Fourier transform of the Hartmanngram located at the distance $z'/p'F$ along the V-axis.

7) The inverse Fourier transform of the result is computed. The phase of the resulting function is proportional to the WFE slopes along Y-axis $\partial \Delta(x,y)/\partial y$. Thus both WFE slopes are determined from a single image acquisition.

8) The final step (not shown on the figure) usually consists in reconstructing the WFE $\Delta(x,y)$ from its estimated slopes $\partial \Delta(x,y)/\partial x$ and $\partial \Delta(x,y)/\partial y$.

### 2.4 Reverse Hartmann Wavefront Sensor (RH-WFS)

Ways of applying the principle of the reverse Hartmann test to an operational WFS were extensively discussed in Ref. [5]. Only the main conclusions are summarized here.

- Achieving quantitative WFE measurements with sufficient accuracy requires optimizing critical parameters $p'$ and $z'$, i.e. the spatial period of the filter and its distance to the camera. For optimal Signal-to-Noise Ratio (SNR) the gain and contrast factors (Eqs. 3b and 3d) should be maximized. Moreover, one also has to minimize the pupil replication effect that is governed by the relative pupil shear criterion $\rho$:

$$\rho = \lambda(F+z')/p'D.$$
(6)

From the here above constraints, the best couple of parameters ($p'$, $z'$) is determined by using a non-linear minimization algorithm.

- For an optimized RH-WFS, numerical simulations in Ref. [5] showed that the absolute WFE reconstruction accuracy ranges from $\lambda/100$ to $\lambda/30$ RMS, depending of the choice of the filter and of other operational parameters. For example better results are obtained when replacing the Hartmann mask with a square Ronchi grid. Working with low aperture numbers $F/D$ and at lower wavelengths $\lambda$ also is more favourable.

- From the practical point of view, the RH-WFS only requires simple hardware, i.e. a grid of pinholes (the Hartmann mask) and the backward gazing camera as depicted in Figure 2. No critical optical components such as microlens array, oscillating pyramidal prism, or achromatic phase plate employed in other types of WFS are needed.

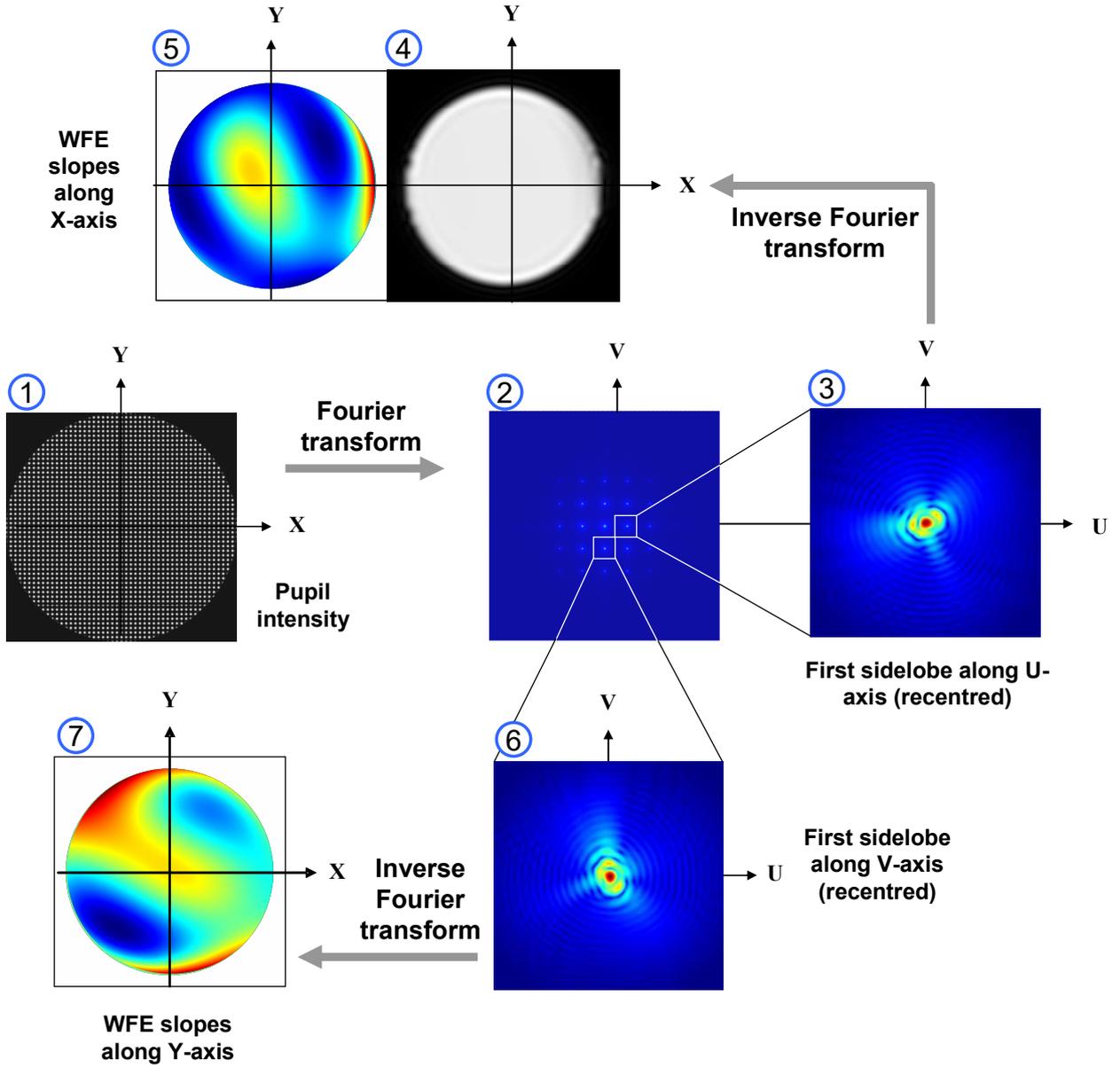

Figure 4: Illustrating the wavefront slopes reconstruction procedure (see text for details).

- Finally, the RH-WFS have the ability to operate over a large spectral bandwidth δλ. In that case the monochromatic contrast $C(\lambda)$ in Eq. 3d shall be replaced with a "polychromatic contrast" $C_{\delta\lambda}(\lambda_0)$ that writes as [5]:

$$C_{\delta\lambda}(\lambda_0) = \mathrm{sinc}\left[\pi \frac{z'(F+z')}{2Fp'^2}\delta\lambda\right]C(\lambda_0), \qquad (7)$$

with sinc($u$) the sine cardinal function sin($u$)/$u$ and $\lambda_0$ the mean wavelength. This criterion can easily be added into the system optimization procedure.

## 3    APPLICATION TO SHACK-HARTMANN WAVEFRONT SENSOR

After having described the theory of the RH-WFS in § 2, this section presents a heuristic application to the case of the now classical SH-WFS. After briefly summarizing the principle (§ 3.1), a global reconstruction procedure of the WFE slopes named SH-IFT (Shack-Hartmann inverse Fourier transform) is presented in § 3.2. The SH-WFS itself is fully modelled in Fourier optics theory as described in § 3.3.

### 3.1    Brief presentation of SH-WFS

Starting from the historical direct Hartmann test sketched in Figure 1-A, R. Shack and B. Platt introduced its modern version in a short note dated 1971 [7]. The principle consists in replacing the Hartmann grid with an array of micro-lenses splitting the optical beam of diameter $D$ in the XY pupil plane into a series of $N \times N$ sub-pupils (see Figure 5). Each microlens focuses light onto a detector array located in the X'Y' image plane. For a given sub-pupil denoted by the indices $m$ and $n$ (both comprised between $-M$ and $+M$ as in § 2.2) the maximal peak intensity[1] is located at a point of Cartesian coordinates $(x'_m, y'_n)$:

$$x'_m = m\,p - f\,\partial\Delta(x,y)/\partial x,$$
$$y'_n = n\,p - f\,\partial\Delta(x,y)/\partial y,$$
(8)

with $p$ the pitch of the micro-lens array equal to $D/N$, and $f$ the focal length of the individual micro-lens. The evident similarity between these relations and those applicable to the RH-WFS (see Eqs. 4) suggests that the same WFE slopes reconstruction procedure as described in § 2.3 could be applied to the intensity distributions recorded on the SH-WFS detector array. This is the scope of the next subsection.

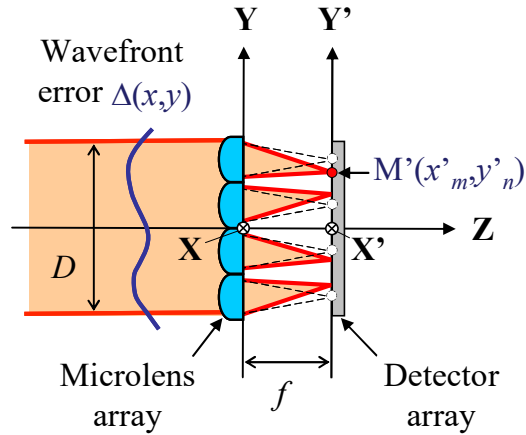

Figure 5: Principle of the SH-WFS.

### 3.2    Global reconstruction of WFE slopes. The SH-IFT method

Comparing the basic SH-WFS relations in Eqs. 8 to Eqs. 4 allows defining an empirical equivalence between all employed quantities and parameters. In particular:

- XY and X'Y' coordinates are exchanged in all analytical relations: for the reverse Hartmann test the so-called Hartmanngram $I(x,y)$ is directly measured in the XY pupil plane (see Figure 2), where the points of maximal intensities are located at Cartesian coordinates ($x_m$, $y_n$). Conversely, the SH-WFS records intensity distributions $I'(x',y')$ in the X'Y' image plane and the maximal intensity peaks are located at coordinates ($x'_m$, $y'_n$).

---

[1] It is well known that SH-WFS generally makes use of centroïding or equivalent algorithms for determining the $(x'_m, y'_n)$ coordinates, instead of localizing the maximal intensity peaks. It may be noticed that the same algorithms could be applied to the Hartmanngrams measured with the RH-WFS as well, thus the similarity remains valid.

- For a SH-WFS the parameters $z'$ and $d''$ in Eqs. 4 loose their significations. The diffracting distance $d''$ is replaced with the focal length $f$ of the micro-lenses and the pitch $p$ of the micro-lens array takes the place of the spatial filter period $p'$. It follows that the gain factor $G$ in Eq. 3b becomes equal to $2\pi f / p$.

- It is also possible to define an equivalent relative pupil shear (as previously defined by Eq. 6) as $\rho = \lambda f/pD$.

- Finally, the Fresnel diffraction parameter $d''$ in Eq. 3c becomes infinite, which means that both monochromatic and polychromatic contrasts in Eqs. 3d and 7 are always maximal and equal to unity.

Therefore the WFE slopes reconstruction procedure applicable to the RH-WFS as defined by Eqs. 5 should be extrapolated to the SH-WFS under the analytical form:

$$\frac{\partial \Delta(x,y)}{\partial x} = \frac{p}{2\pi f} Arg\{FT^{-1}[B_{1/p}(u,v)FT[I'(x',y')](u-1/p,v)]\}$$
$$\frac{\partial \Delta(x,y)}{\partial y} = \frac{p}{2\pi f} Arg\{FT^{-1}[B_{1/p}(u,v)FT[I'(x',y')](u,v-1/p)]\}$$
(9)

Unlike the conventional slopes reconstruction process that makes use of local centroïd determination algorithms or equivalent local phase-fitting algorithms in the Fourier plane, it must be pointed out that the SH-IFT procedure is global: here the recorded intensity distribution $I'(x',y')$ is considered as a single Hartmanngram, from which slopes errors are retrieved using the double Fourier transform algorithm illustrated in Figure 4.

### 3.3 Fourier optics model of the SH-WFS

A Fourier optics model of the SH-WFS has finally been developed in order to test the SH-IFT slopes reconstruction procedure described in the previous subsection. Unlike the RH-WFS which requires Fresnel diffraction theory, the SH-WFS model involves Fraunhofer diffraction only and is schematically illustrated in Figure 6. Here the micro-lens array is represented as a phase screen map noted $\Gamma(x,y)$ and located in the pupil plane of the optical system, where it is splitted into $N$ x $N$ sub-areas. Each of them defines the contour of an individual micro-lens located at the Cartesian coordinates ($mp$, $np$) and is carrying a linear phase ramp proportional to those coordinates. Mathematically, $\Gamma(x,y)$ writes as:

$$\Gamma(x,y) = \frac{2\pi}{\lambda f} \sum_{m=-M}^{+M} \sum_{n=-M}^{+M} S_p(x-mp, y-np)[mp(x-mp) + np(y-np)],$$
(10)

where $S_p(x,y)$ is a "boxcar" function equal to unity inside a square of side $p$ and to zero outside of it. Modeling the recorded intensities $I'(x',y')$ is thus carried out as follows:

1) A wavefront $\Delta(x,y)$ is computed from a given set of Zernike polynomial coefficients. Figure 6-1 shows a false-colour representation of such an input WFE.

2) The phase function $\Gamma(x,y)$ is evaluated from Eq. 10. The false-colour view of $\Gamma(x,y)$ and its profile along the X-axis are shown in Figure 6-2.

3) The total phase at the output of the micro-lens array is computed as $2\pi\Delta(x,y)/\lambda + \Gamma(x,y)$ and illustrated in Figure 6-3.

4) Finally, determining the intensity measured by the detector array consists in a classical Fraunhofer diffraction calculation, firstly computing the Fourier transform of the complex amplitude at the exit of the micro-lens array, then taking the square modulus of the result:

$$I'(x',y') = |FT[B_D(x,y)\exp[2i\pi\Delta(x,y)/\lambda + i\Gamma(x,y)]]|^2,$$
(11)

with $i$ the complex square root of $-1$. An example of resulting intensity distribution $I'(x',y')$ is shown in Figure 6-4.

The next steps of the procedure consist in reconstructing the WFE slopes $\partial\Delta(x,y)/\partial x$ and $\partial\Delta(x,y)/\partial y$ from the Hartmanngram $I'(x',y')$. These operations are strictly similar to those described in steps 2-7 of section 2.3, which are not repeated here. The graphic illustration in Figure 4 also remains valid, the sole difference being that the input Hartmanngram (i.e. $I(x,y)$ in Figure 4-1 is replaced with $I'(x',y')$ in Figure 6-4.

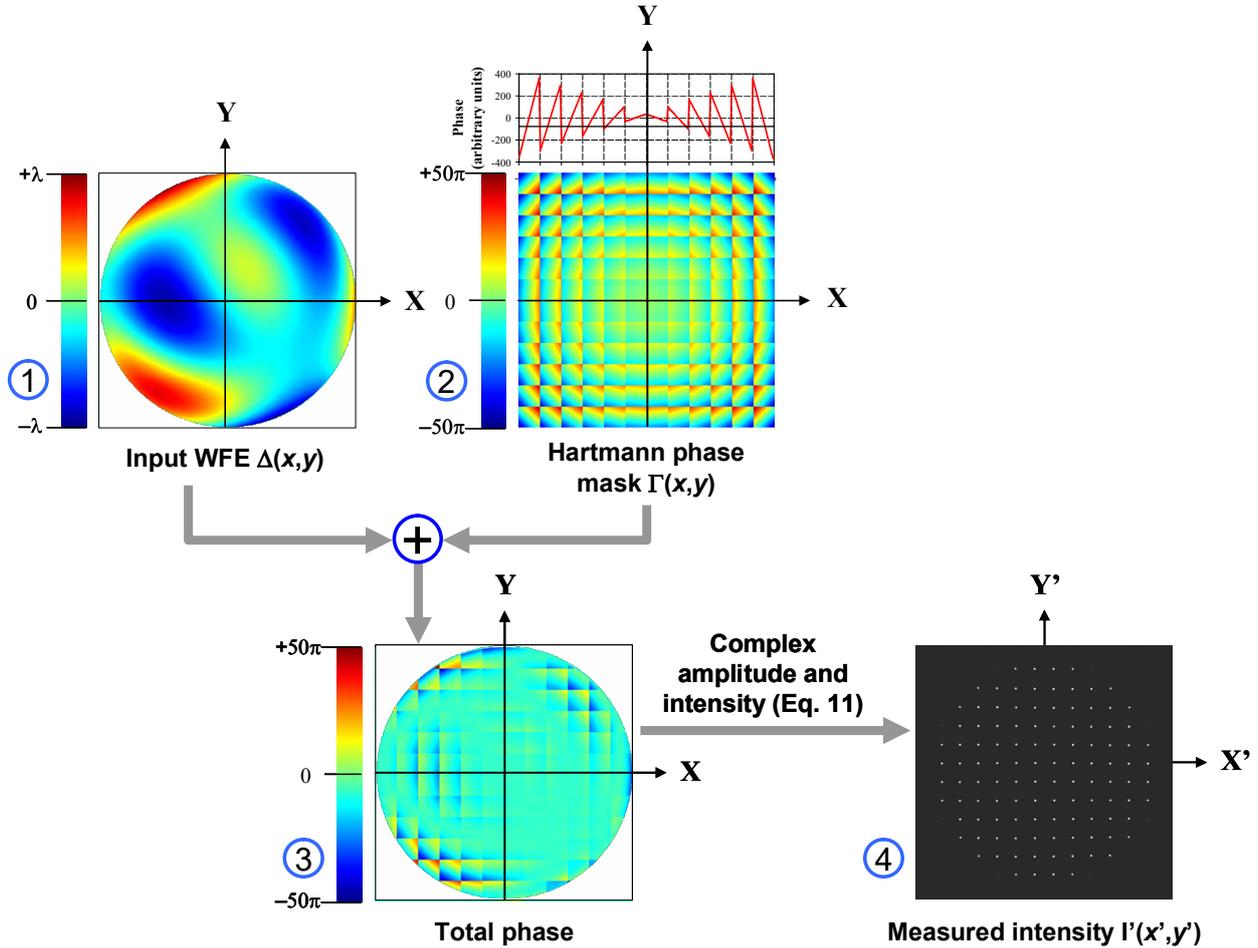

Figure 6: Fourier optics model of the SH-WFS (see text for details).

## 4 NUMERICAL SIMULATIONS

### 4.1 Genera approach and simulation parameters

A fair comparison between the SH-WFS and RH-WFS is only possible by using the same family of optical models for both devices. It includes the simulations of the intensity distributions recorded by the detector array, on the one hand, and the WFE slopes reconstruction procedure, on the other hand. For that purpose there exists a vast choice of geometrical or Fourier optics models, as summarized in Table 1. Here it has been decided to do not mix geometrical and physical optics theories, and to stay in the general framework of Fourier optics. Therefore,

- Numerical simulations of the Hartmanngrams $I(x,y)$ produced with the RH-WFS are carried out using the Fresnel diffraction model described in Ref. [5], section 4. The slopes reconstruction procedure is the same as in Ref. [5] and as summarized in subsection 2.2 of the present paper,

- Numerical simulations of the intensities $I'(x',y')$ recorded by the SH-WFS make use of the Fraunhofer diffraction model described in § 3.3. Slopes reconstructions are performed using the SH-IFT method (§ 3.2).

Table 1: Different optical models usable for numerical simulations.

| Theory | Square Ronchi / Reverse Hartmann | | Schack-Hartmann | |
|---|---|---|---|---|
| | Image simulation model | WFE slopes reconstruction procedure | Image simulation model | WFE slopes reconstruction procedure |
| Geometrical optics | Ray-tracing model | Centroid determination model | Ray-tracing model | Centroid determination model |
| Fourier optics | Fresnel diffraction model [5] | Fourier transform method [6] | Fraunhofer diffraction model (§ 3.3) | SHIFT method (§ 3.2) |

Let us now consider an optical system whose optical aberrations are to be measured successively by using the square Ronchi and Hartmann tests (i.e. two variants of a RH-WFS), then with a conventional SH-WFS. The general parameters of the optical system and of its test setup are given in Table 2. For each of those measurement configurations the following cases are studied:

- Two different pupil sampling are simulated, respectively equal to $N \times N = 33 \times 33$ and $65 \times 65$. For the RH-WFS it implies that the period of the spatial filter $p'$ and its distance to the observing camera $z'$ have been previously optimized in order to match the required pupil sampling while preserving the measurement accuracy (see § 2.4 and Ref. [5], section 3.A for more details about the optimization procedure). The resulting values of $p'$ and $z'$ are thus indicated in Table 2. For the SH-WFS the pupil sampling is simply equal to the number of micro-lenses forming the array.

- The WFE to be measured $\Delta(x,y)$ is generated from the first 16 Zernike polynomials whose amplitudes are selected randomly. It is then rescaled to have its Peak-to-Valley (PTV) value around $4\lambda$, and differentiated numerically in order to define the reference WFE slopes $\partial\Delta(x,y)/\partial x$ and $\partial\Delta(x,y)/\partial y$ to be measured by the WFS. False color views of the WFE and slopes maps are depicted in Figure 7-1. The PTV and RMS values of $\partial\Delta(x,y)/\partial x$ and $\partial\Delta(x,y)/\partial y$ are given in the left column of Table 3.

It must be noted that in the SH-WFS case the focal length of the micro-lenses is assumed to be equal to the focal length of the tested optical system, because the simplified optical model presented in § 3.3 does not take into account unavoidable scale changes between the diameters of the pupil and of real micro-lens arrays. For all cases, Table 2 also indicates the achieved values of the relative pupil shear $\rho$, gain $G$, and monochromatic contrast $C(\lambda)$ where applicable.

## 4.2 Numerical results

The main results of the numerical simulations are summarized in Table 3, giving for each test configuration the PTV and RMS values of the measured slopes along both X and Y axes (central column), and of their difference with respect to the reference slopes maps (right column). They are also illustrated by the false-color views of Figure 7 that show, from top to bottom:
- in Figure 7-2, the measured slopes along the X and Y axes and the resulting errors maps obtained with the square Ronchi grating for a 33 x 33 pupil sampling,
- in Figure 7-3, same illustrations for the square Ronchi test with a 65 x 65 pupil sampling,

- in Figure 7-4, same illustrations for the reverse Hartmann test, where the spatial filter is a simple grid of 33 x 33 pinholes,
- in Figure 7-5, same illustrations for the reverse Hartmann test with a grid of 65 x 65 pinholes,
- in Figure 7-6, same illustrations for a Shack-Hartmann measurement using the SH-IFT method with a 33 x 33 micro-lens array,
- in Figure 7-7, same illustrations for the Shack-Hartmann with a 65 x 65 micro-lens array.

Table 2: System parameters for RH-WFS and SH-IFT numerical simulations.

| Parameters | Symbol / Formula | Pupil sampling N x N | | Unit |
|---|---|---|---|---|
| | | 33 x 33 | 65 x 65 | |
| *General / Tested optical system* | | | | |
| Reference wavelength | $\lambda$ | 0.6 | 0.6 | µm |
| Focal length | $F$ | 1 | 1 | m |
| Diameter | $D$ | 0.3 | 0.3 | m |
| Aperture number | $F/D$ | 3.3 | 3.3 | – |
| *Square Ronchi / Reverse Hartmann test* | | | | |
| Image to filter distance | $z' = O_1O'$ | -0.281 | -0.490 | m |
| Pupil to filter relative distance | $\zeta = (F+z')/F$ | 0.719 | 0.510 | – |
| Filter period | $p'$ | 2.64 | 2.30 | mm |
| Filter spatial frequency | $v' = 1/p'$ | 0.38 | 0.44 | mm$^{-1}$ |
| Relative pupil shear | See Eq. 6 | 0.05 | 0.04 | % |
| Gain | See Eq. 3b | 1.7E+03 | 1.4E+03 | – |
| Contrast (monochromatic) | See Eq. 3d | 0.999 | 0.996 | – |
| *Shack-Hartmann with SH-IFT method* | | | | |
| Microlens array pitch / Microlens width | $p = D/N$ | 9.09 | 4.62 | mm |
| Microlenses focal length | $f$ | 1000 | 1000 | mm |
| Relative pupil shear | $\rho = \lambda f/pD$ | 0.022 | 0.043 | % |
| Gain | $G = 2\pi f/p$ | 6.9E+02 | 1.4E+03 | – |

Examining the results in Table 3 and the plots of the error maps in Figure 7 allows determining some clear tendencies:

- The best results in terms of measurement accuracy are obtained using either a RH-WFS equipped with a square Ronchi grating, or with a SH-WFS operating with the SH-IFT method. In both cases, the relative slopes measurement errors are found to be below 10% at a pupil sampling of 65 x 65. This is very satisfactory since from Ref. [5], section 4, WFE measurement errors are typically 3 or 4 lower than the slopes measurement errors after WFE reconstruction. Conversely, operating with a RH-WFS equipped with a simple Hartmann grid tends to decrease its potential measurement accuracy.

- For all WFS types and whatever spatial filter is employed, more accurate results are always obtained with the 65 x 65 pupil sampling, which may not look very surprising. It can be extrapolated that better results are achievable using higher and higher pupil sampling. From a practical point of view, that point pleads in favor of the RH-WFS vs. SH-WFS, since the manufacturing of high pupil sampling spatial filters is probably much easier than their equivalent micro-lens arrays.

- Examining false-color views of the slopes errors maps of the reverse Hartmann WFS confirms that the most important measurement errors are located near the pupil rim. This effect has already been noticed in Ref. [5] and could eventually be mitigated with the help of spatial filtering algorithms in the Fourier plane (not implemented in the present study). On the contrary, the main limitations of the SH-IFT method seem to originate from the pitch of the micro-lens array that is clearly revealed in Figure 7-6, and to a lesser extent in Figure 7-7 at a higher pupil sampling.

From a pure programmer's point of view, it has to be noted that the Fourier optics model of the SH-WFS presented in § 3.3 requires extremely large variable arrays in order to obtain a sufficient sampling at the entrance face of each individual micro-lens. For practical reasons we were limited to arrays of dimensions 32767 x 32767 (including a necessary zero-padding of the pupil plane to avoid aliasing effect), which may limit the accuracy of these numerical simulations. A side conclusion of this study should be the need of using more powerful computers in order to modeling the SH-WFS in a pure Fourier optics framework.

Table 3: Numerical results of RH-WFS and SH-IFT simulations.

|  | Original WFE slopes | Measured slopes | Slopes difference | Relative error (%) | Original WFE slopes | Measured slopes | Slopes difference | Relative error (%) |  |
|---|---|---|---|---|---|---|---|---|---|
| **Square Ronchi test** | Pupil sampling 33 x 33 | | | | Pupil sampling 65 x 65 | | | | |
| X-slopes (mrad) | 0.052 0.010 | 0.070 0.010 | 0.059 0.003 | 114 26 | 0.052 0.010 | 0.057 0.010 | 0.017 0.001 | 32 9 | PTV RMS |
| Y-slopes (mrad) | 0.060 0.012 | 0.099 0.013 | 0.064 0.003 | 107 25 | 0.060 0.012 | 0.068 0.012 | 0.018 0.001 | 30 8 | PTV RMS |
| **Reverse Hartmann** | Pupil sampling 33 x 33 | | | | Pupil sampling 65 x 65 | | | | |
| X-slopes (mrad) | 0.052 0.010 | 0.061 0.011 | 0.043 0.005 | 83 55 | 0.052 0.010 | 0.056 0.010 | 0.011 0.002 | 21 17 | PTV RMS |
| Y-slopes (mrad) | 0.060 0.012 | 0.081 0.011 | 0.046 0.006 | 76 48 | 0.060 0.012 | 0.059 0.011 | 0.015 0.002 | 25 19 | PTV RMS |
| **SH-IFT method** | Pupil sampling 33 x 33 | | | | Pupil sampling 65 x 65 | | | | |
| X-slopes (mrad) | 0.051 0.010 | 0.054 0.010 | 0.009 0.001 | 17 13 | 0.051 0.010 | 0.052 0.010 | 0.006 0.001 | 12 8 | PTV RMS |
| Y-slopes (mrad) | 0.060 0.012 | 0.063 0.012 | 0.009 0.002 | 15 14 | 0.060 0.012 | 0.059 0.011 | 0.006 0.001 | 10 8 | PTV RMS |

## 5 CONCLUSION

In this communication were presented pure Fourier optics theories of two important classes of WFS based on the historical "direct" Hartmann test (that gave birth to moderns SH-WFS) and on its "reversed" version where the locations of the pupil and image planes are exchanged. The latter may give rise to a new generation of wavefront sensors, named RH-WFS. After summarizing its theoretical model that involves Fresnel diffraction, we described its specific WFE reconstruction procedure based on a double Fourier transform algorithm. A similar approach was followed for classical SH-WFS, allowing a comparison of both concepts in the same Fourier optics framework. Numerical simulations were carried out in order to determining the achievable measurement accuracy of three different test configurations, including the SH-WFS and a RH-WFS equipped with either square Ronchi or pinholes grid spatial filters. The numerical results show equivalent performance in terms of measurement accuracy of the reversed concept when employing a square Ronchi grating, and of the direct one using the SH-IFT method. From a practical point of view, it should be noted however that the RH-WFS incorporates no critical optical component such as the high sampling micro-lens arrays required by a SH-WFS.

Another important finding of the study is the apparent good performance of the SH-IFT method when applied to the raw images recorded by the SH-WFS. It may be due to the fact that the whole intensity distributions are processed globally by the SH-IFT algorithm, without reducing the information to the sole coordinates of the spots centroids as is usually done. Future work is desirable to confirm that hypothesis, this time applying similar WFE slopes reconstruction methods based on classical centroiding algorithms to both types of WFS.

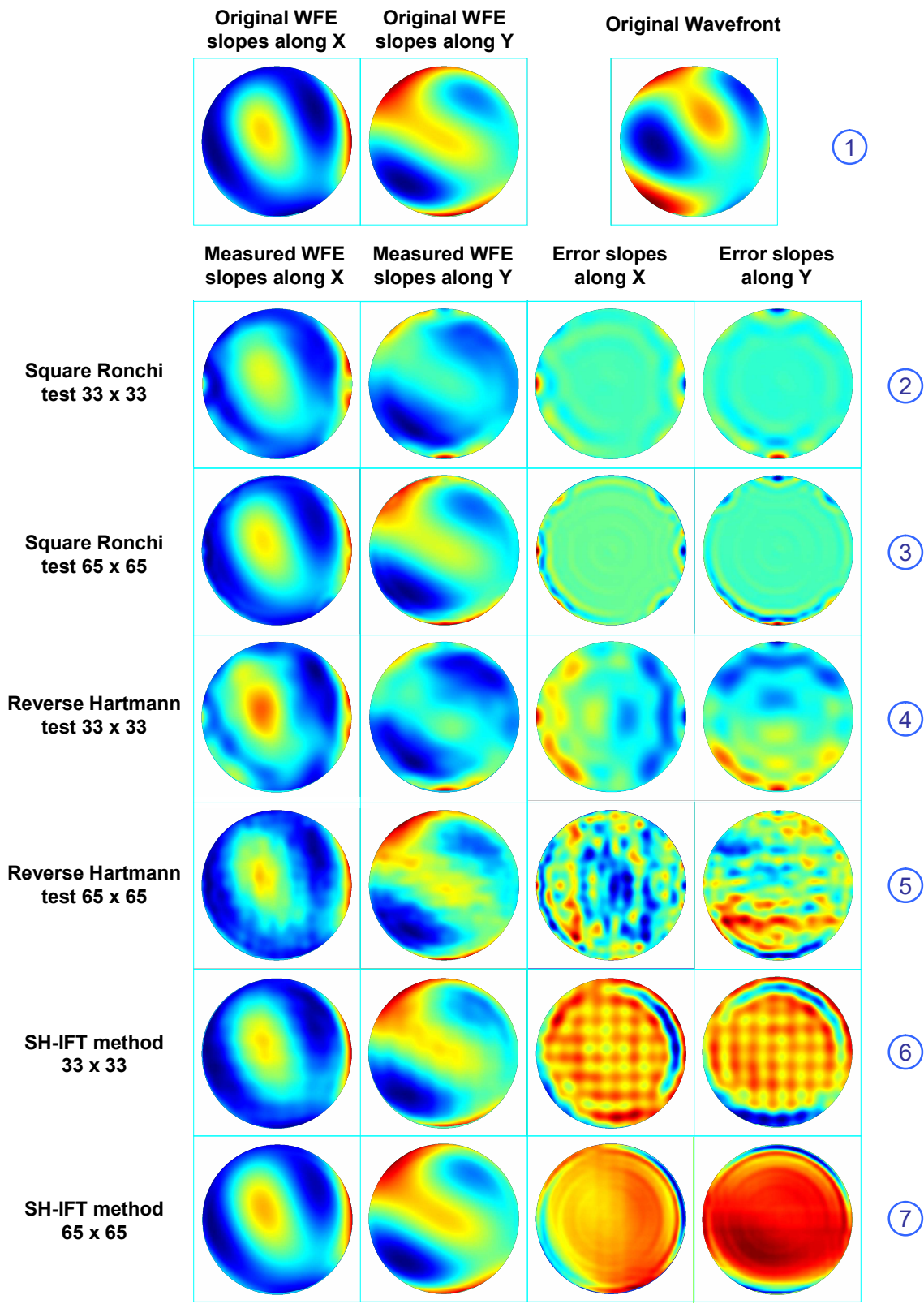

Figure 7: False color views of numerical simulations.